\documentclass[aps,pre,twocolumn,groupedaddress]{revtex4}
\usepackage{graphicx}
\def\be{\begin{equation}}
\def\ee{\end{equation}}
\begin{document}
\draft

\title{Wave breaking and particle jets in intense inhomogeneous 
charged beams}
\author{Felipe B. Rizzato\footnote{rizzato@if.ufrgs.br}, Renato Pakter\footnote{pakter@if.ufrgs.br}, 
and Yan Levin\footnote{levin@if.ufrgs}}
\address{Instituto de F\'{\i}sica,
Universidade Federal do Rio Grande do Sul\\
Caixa Postal 15051, 91501-970, Porto Alegre, RS, Brazil
}
%\nodate{\nodate}
\begin{abstract}
%.......................................
This work analyzes the dynamics of inhomogeneous, magnetically focused high-intensity beams of charged 
particles. While for homogeneous beams the whole system oscillates with a single frequency, any 
inhomogeneity leads to propagating transverse density waves which eventually result in a singular density 
build up, causing wave breaking and jet formation. The theory presented in this paper 
allows to analytically calculate the time at which the wave breaking takes place. It also gives a good estimate 
of the time necessary for the beam to relax into the final stationary state consisting of a cold core surrounded 
by a halo of highly energetic particles. 
%.......................................
\end{abstract}
\maketitle
%.......................................
%
%\newpage
%\section{Introduction}
%
It is well known that magnetically focused beams of charged particles can relax from 
non-stationary into stationary flows with the associated particle evaporation~\cite{reiser91}. This is
the case for homogeneous beams with initially mismatched envelopes flowing along the
magnetic symmetry axis of the focusing system. 
Gluckstern \cite{gluck94} showed that initial oscillations of mismatched beams induce formation
of large scale resonant islands \cite{pak94,pak97} beyond the beam border: beam
particles are captured by the resonant islands resulting in emittance growth and relaxation.
A closely related question concerns the mechanism of beam relaxation and the associated emittance growth 
when the beam is not homogeneous. On general grounds of energy conservation 
one again concludes that beam relaxation takes place as the coherent fluctuations of beam inhomogeneities 
%AQUI
are converted into microscopic kinetic and field energy \cite{gluc70,reiser91,lund98,bern99}.
However, unlike  in the former case for which the specific resonant mechanism is 
well understood, for inhomogeneous systems a more detailed description of the 
processes involved must still be explored. This is the goal of the present letter.

The interest surrounding a better understanding of the dynamics of inhomogeneous beams is due to the 
fact that one can hardly design experimental devices capable of generating fully matched beams at 
the entrance of a transport system \cite{lund05}. While the azimuthal symmetry with respect to the beam axis is 
feasible in the case of focusing solenoidal magnetic fields, envelope matching --- to avoid the radial oscillations ---
is significantly harder to achieve,  while a perfect homogeneity is practically impossible.  

Given all these facts, the purpose of the present work is to investigate the mechanisms
leading to the decay of density inhomogeneities 
as the system relaxes into its final stationary state.  We find that the relaxation 
comes about as a consequence of 
breaking of density waves followed by ejection of fast particle jets. 
Jets are formed by particles moving in-phase with the macroscopic density fluctuations. They  
draw their energy from the propagating wave fronts and convert it into 
microscopic kinetic energy. This process is 
very similar to the breaking of gravitational surface waves.  The jet can then be compared to
a broken crest of a gravitational wave surfing down the wave front.
%These in-phase particles cannot overcome the potential 
%energy of wave fields, and this is equivalent to what happens with wave breaking of gravitational 
%waves on water. Several details differ, including the fact that we have transverse 
%breaking here \cite{bu06}, but the general idea is similar.
%Jets draw energy from macroscopic field fluctuations through particles moving 
%in-phase with these fluctuations, and energy is thereafter delivered to 
We stress that the wave breaking mechanism analyzed in this letter is very different from the Gluckstern 
resonances which were found to be the driving force behind the emittance growth in transversely oscillating 
homogeneous particle beams.  For strongly inhomogeneous beams, we find that it is the wave breaking and jet
production which are the primary mechanisms responsible for the beam relaxation.

%Since breaking can be dominant under certain general circumstances to be described later, an adequate 
%analysis is in order.

%Considering the current relevance of intense beams, 
We consider solenoidal focusing of 
space-charge dominated beams propagating along the transport axis, defined as the $z$ axis, of our reference frame. 
The beam is initially cold with vanishing emittance, and is azimuthally
symmetric around the $z$ axis. Since the number of 
constituent particles is very large, the beam dynamics is governed by 
the azimuthally equation and collective effects are dominant. Prior to the appearance of density singularities, 
the original Vlasov formalism can be simplified to a cold fluid description for which Lagrangian coordinates are 
particularly appropriate. 
In these coordinates, the transverse radial position $r$ of a beam element is governed by 
\cite{ikegami97,davidson01,pak01,bru95}
\begin{equation}
r'' = - \kappa\,r + {Q(r_0) \over r}.
\label{equa1}
\end{equation}
The prime indicates derivative with respect to the longitudinal $z$ coordinate which for
convenience we shall also refer to as ``time'', and angular momentum in the Larmor frame is taken to be zero 
for each particle.
%
%AQUI 
The focusing factor is 
$\kappa \equiv (q B /2 \gamma m \beta c^2)^2$, where $B$ is the axial, constant, focusing 
magnetic field; $Q(r_0)=K N(r_0)/N_t$,  
is the measure of the charge contained 
between the origin at $r=0$ and the initial position $r(z=0)=r_0$,  $N_t$ is the 
total number of beam particles per unit axial length, 
$N(r_0)$ is the number of particles up to $r_0$, and $K=N_t q^2 /\gamma^3 m \beta^2 c^2$ is the beam perveance.
$q$ and $m$ denote the beam particle charge and mass, respectively; $\gamma = (1-\beta^2)^{-1/2}$ is the 
relativistic factor where $\beta = v_z/c$ and $v_z$ is the constant axial beam velocity and $c$ is the speed of light. 
Note that
$r_0$ is in fact the Lagrange coordinate of the fluid element \cite{morr98} which means that as long as 
the fluid description remains valid, 
the amount of charge seen by the fluid element inside the region $0<r\le r(z)$ 
remains unaltered at $Q(r_0)$, independent of time $z$. 
This is of fundamental importance since from the Gauss law this is the charge that exerts the force on the fluid 
element. In this letter we will consider the beams starting from a static initial 
condition, $r'(0)=0$. The formal solution to the fluid dynamics Eq. (\ref{equa1}) is 
$r=r(z,r_0)$.  This can be calculated explicitly using the Lindstedt-Poincar\'e 
perturbation theory.  For small amplitude fluctuations around the stable equilibrium 
$r_{eq}(r_0)=\sqrt{Q(r_0)/\kappa}$ we obtain
\begin{eqnarray}
r(r_0,z)=r_{eq}\, \big\{1 + A/r_{eq} \cos(\omega z) +(1/3)\>(A/r_{eq})^2 \> \times \cr 
[2+ \cos(\omega z)] \sin^2(\omega z/2)+ {\cal O} [(A/r_{eq})^3]\big\}\;,
\label{equa1p5}
\end{eqnarray}
where $A(r_0)=r_0-r_{eq}$ is the amplitude of oscillations, 
$\omega(r_0) = \omega_0+\sqrt{\kappa}\,A^2/(6 \sqrt{2}\,r_{eq}^2)$ is the renormalized
$r_0$-dependent frequency, and $\omega_0 = \sqrt{2 \kappa}$ is the unperturbed frequency.  We 
stress that as long as the fluid picture applies, all the information about the 
temporal evolution of the beam is contained
in Eq.~(\ref{equa1p5}).  For example, the time evolution of the beam density
can be obtained as follows. For a beam of initial cross sectional density $n_0(r)$, 
the amount of charge $\delta Q$ between two 
concentric circles of radii $r_0$ and $r_0+\delta r_0$ ($\delta r_0$ small) 
is $\delta Q = 2 \pi r_0 \delta r_0 (K/N_t) n_0 (r_0)$. 
Since this charge is conserved,  
$\delta Q = 2 \pi r \delta r  \> (K/N_t)\, n\left(r \right) = 2 \pi r 
( \partial r / \partial r_0 ) (K/N_t) \, n(r) \delta r_0$, the transverse beam density at any future time $z$ is,
therefore,
\begin{equation}
n(r) = n_0(r_0) \, (r_0 / r) \,\left( \partial r / \partial r_0 \right)^{-1}\big|_{r_0 = r_0 (r,z)}.
\label{equa2}
\end{equation}
For a given position $r$ and an axial coordinate $z$, the initial position 
$r_0$ of a beam element can be uniquely determined   
as the inverse function $r_0 = r_0 (r,z)$ of Eq. (\ref{equa1p5}). 
This ceases to be the case if $\partial r / \partial r_0 \rightarrow 0$ and $r_0 (r,z)$ becomes
multivalued. If this happens, the density will diverge and the fluid picture will break down. 
All these features, if present, would be indicative of a wave breaking phenomenon. 
%When this happens some particles will become 
%entrapped by the waves, stagnation points will form, and the density undergoes an abrupt growth. 
%Wave breaking saturates as the smooth fluid breaks up into several kinetic jets of energetic particles. 
Needless to say that presence of wave breaking in 
charged particle beams would be of considerable interest and practical importance. Breaking might be responsible for 
conversion of energy from macroscopic fluid modes into microscopic kinetic activity. 

\begin{figure} [ht]
\begin{center}
\includegraphics[scale=0.25,angle=270]{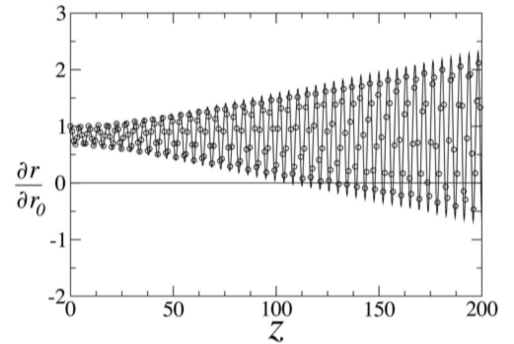}% Here is how to import
%EPS art
\end{center}
\caption{Time evolution of the compressibility factor $\partial r/\partial r_0$ for $\chi=0.6$ and $r_0=0.4 r_b$. 
$\partial r/\partial r_0=0$ indicates singular density. Symbols represent the compressibility obtained from the
numerical solution of Eq. (\ref{equa1}), 
whereas the solid line is derived using the perturbative solution, Eq. (\ref{equa1p5}). $z$ is measured in units of 
$\kappa^{-1/2}$. Position $r_0=0.4 \,  r_b$ is chosen
because this is the approximate Lagrange coordinate at which density is first found to diverge for 
$\chi =0.6$.\label{fig1.eps}}
\end{figure}

We start our analysis by considering the compressibility factor 
$\partial r / \partial r_0$, which can be obtained exactly by 
numerically integrating two nearby trajectories of
Eq. (\ref{equa1}) or approximately by differentiating Eq. (\ref{equa1p5}). 
%Given the initial beam profile it is then possible to 
%evaluate the density $n$ using Eq. (\ref{equa2}).  
To be specific we write the initial cross sectional beam 
density at $z=0$ in a general parabolic form
$n_0(r_0) =\rho_h + \chi \rho_i(r_0) $, where the inhomogeneity parameter $0\le\chi\le1$, $\rho_h \equiv N_t / (\pi r_b^2)$,
and $\rho_i(r_0) \equiv \rho_h (2 r_0^2 / r_b^2-1)$.  $r_b$ is the beam radius. 
Note that the integral of the inhomogeneous contribution $\rho_i(r_0)$ is zero.
To suppress the effects arising from the pure envelope 
oscillations --- Gluckstern resonances --- we fix $r_b=\sqrt{K/\kappa}$, 
so that the beam radius is unaltered for as long as the fluid picture,
Eq. (\ref{equa1}), remains valid. 
%In contrast to the previous works in which the density profile was kept
%constant while the envelope oscillated \cite{chi95}, our envelope size is fixed and while the density profile 
%evolves with time. 
We have also performed calculations for 
rms matched beams and find that they behave qualitatively the same way. 
Fig. \ref{fig1.eps} shows the typical time evolution of the compressibility $\partial r / \partial r_0$ obtained both 
numerically using Eq. (\ref{equa1}) and analytically using (\ref{equa1p5}). 
Since the amplitudes of oscillations of all fluid elements about the points of their 
equilibria are small, $|A/r_{eq}| \ll 1$, even for large 
values of $\chi$,  the agreement between the numerical and the perturbative solution 
is found to be very good, see Fig. \ref{fig1.eps}.
The compressibility factor exhibits 
a fast oscillatory motion accompanied by a slow secular growth.  This means that given enough time, the 
compressibility will always become zero for any finite value of $\chi$, resulting in a 
density divergence.  
%This is very suggestive of a wave breaking pehnomenon with all the 
%associated features. 

To further explore the significance of the diverging density, 
we have performed fully self-consistent N-particle simulations. 
For a system in which particles interact by an infinite range unscreened Coulomb potential,  
the time of collision diverges and the mean-field Vlasov description becomes exact~\cite{brhe}.  
Thus, in order to simulate the $N_t \rightarrow \infty$ limit --- in which the thermalization
time due to binary collisions is infinite --- each particle can be taken to interact 
only with the mean-field produced by all the other particles.  
Taking advantage of the azimuthal beam symmetry and the Gauss law, 
a particle located at a position $r$ experiences a 
field generated by the particles within 
a circle of radius $r$ \cite{ikegami97,levin02}. We stress
that in the $N_t \rightarrow \infty$ limit this is exact at any 
finite time scale. Within the simulation, 
the trajectory of each particle is, therefore, also governed by
Eq. (\ref{equa1}) --- unlike the fluid elements, however, particles are allowed to bypass one another.
This method avoids the thermalization effects associated with the binary collisions and significantly 
speeds up the simulations.   
It allows us to accurately simulate the collective effects dominant for all time scales 
when $N_t \rightarrow \infty$ even with a relatively small number of particles, $N_t \approx 10000$.

\begin{figure} [h]
\includegraphics[height=11.0cm]{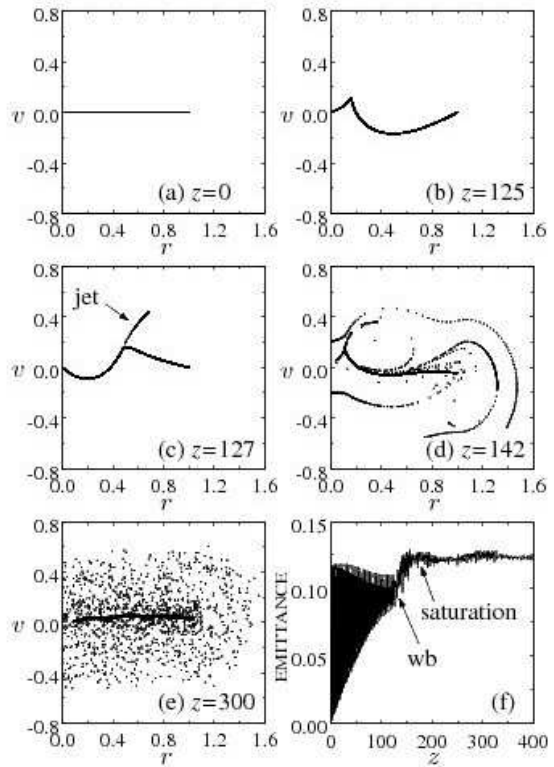}% Here is how to import
%EPS art
\caption{Time evolution in the phase-space $(r,v)$ for $\chi=0.6$. $r$ is measured in units of 
$(K/\kappa)^{1/2}$, $v$ in units of $K^{1/2}$, and $z$ in units of $\kappa^{-1/2}$: 
panel (a) -- initial condition at $z=0$; panel (b) -- after 
many wave cycles, but just before wave breaking; panel (c) -- just after wave breaking,  
an azimuthal jet is expanding over the phase-space; panel (d) -- while the first jet moves in the
phase space new jets are being ejected; panel (e) -- a phase portrait of a relaxed state; 
panel (f) --  emittance growth leading to a final relaxation.  Note that emittance
saturates soon after the wave breaking (wb).\label{fig2.eps}}
\end{figure}

In Fig. \ref{fig2.eps} we display the particle phase-space $(r,v \equiv r')$ for  $\chi = 0.6$. 
The first panel (a) shows the initial distribution at $z=0$ --- all particles are still. 
In panel (b), after various propagating wave cycles, the system is about to build up an infinite density;
velocity is still a single valued function of the space coordinate 
but a singularity (cusp) is forming. The third panel (c) shows the system at a
time slightly larger than the first wave breaking time. Velocity 
ceases to be a single valued function of the space coordinate --- while 
some of the particles go through the wave others do not.  This latter class of particles is accelerated by the 
wave front and forms a thin azimuthally symmetric jet or finger seen in the figure. High energy 
jet particles can reach far outside the beam core and may be very 
detrimental to the beam transport.
The process shown in panel (c) repeats itself many
times, see panel (d), as the system evolves toward a final stationary state and the previously unoccupied 
extensions of the phase-space are gradually filled with particles whose velocities are considerably larger 
than velocities in the beam core.  After some time a stationary state is reached in which 
the beam separates into a cold dense core and a hot and extended halo of ejected 
particles, panel (e). Time evolution of the 
emittance~\cite{davidson01} $\varepsilon \equiv 2\,\sqrt{<r^2><v^2>-<r\,v>^2}$, where $<>$ 
denotes an average over particles, is shown in the last panel (f). At the wave breaking emittance suffers a sharp
rise, followed by a rapid relaxation to the final stationary state 
in which large amplitude fluctuations subside. 
%AQUI2
We note that beam relaxation is closely connected to phase-space filamentation. 
Phase-space filamentation takes place after particles are ejected from the beam core by surfing 
on the charge density waves. Once outside the core, particles experience the time dependent nonlinear 
forces and undergo all the complicated mixing dynamics with subsequent filamentation. This leads to 
final irreversible emittance growth. The directed emittance growth seen prior to relaxation is reversible. 
%======
%
Since the beam radius $r_b$ is matched --- envelope oscillations are small --- 
the contribution of the  Gluckstern resonant mechanism  to the emittance growth and beam relaxation
is, at most, marginal. The dominant mechanism is the 
singular build up of density followed by the wave breaking and jet production.
%: fields grow until particle trapping takes place, particles 
%are subsequently accelerated, emittance grows, and breaking is thus saturated. The process repeats itself a number of 
%times until the dynamics approaches the final stationary state. 
The time of the first wave breaking depicted in the panel 
(c) of Fig. \ref{fig2.eps} agrees well with the time when the compressibility factor 
$\partial r/\partial r_0$ obtained from the Lagrange fluid 
equations goes to zero,  Fig. \ref{fig1.eps}.  Our next goal is then to precisely calculate the
instant at which the wave breaking takes place.  

We first note that for an inhomogeneous density profile, 
each fluid element
oscillates with a different frequency --- rigid oscillations are possible only when 
the density profile across the beam cross section is homogeneous. Thus, nearby fluid 
elements will oscillate around their points of equilibria, 
slowly moving out-of-phase. This motion results in transverse density waves propagating
across the beam.  At some point, however, two nearby fluid elements
will overlap one another leading to a singular build up of density.  When this happens 
the fluid picture will lose its validity and will have to be replaced by the full kinetic
description given in terms of the Vlasov equation.
The wave breaking occurs when the
separation between any two fluid elements vanishes, 
$r(r_0+\delta r_0,z)-r(r_0,z) \rightarrow 0$, for some value of 
$r_0$.  This is precisely equivalent to our condition for the appearance of a singular density, 
$\partial r/\partial r_0 \rightarrow 0$.  Considering  only the term linear in amplitude of 
Eq.~(\ref{equa1p5}), we see that $\delta r=\delta r_{eq} +  \cos(\omega z) \delta A-A z \sin(\omega z) \delta \omega$. 
Neglecting the purely oscillatory term, as compared to the secular one, the  time of 
breaking is found to be
\begin{equation}
z_{wb} \approx \min_{r_0} \left|{1 \over 2 \sqrt{Q}} {\partial Q / \partial r_0 \over A \partial \omega/\partial r_0}\right|.
\label{equa7}
\end{equation}
As expected, the breaking will always occur whenever $\partial \omega / \partial r_0 \neq 0$.  
This is the case for all inhomogeneous particle beams. 
%Contiguous fluid elements, even if initially oscillating in-phase, eventually reach each other generating singular densities. 
Unlike other systems in which one must have strong enough electric fields 
\cite{rosen02}, 
here any sort of inhomogeneity leads to the wave breaking --- one just has to wait sufficiently long. 
As soon as the 
wave breaking takes place,  particles with the same velocity as the density wave will be captured by the wave and 
surf down its front 
gaining kinetic energy (Fig. \ref{fig2.eps}). Since at the wave breaking position $r_0/r_b<1$, 
minimization in Eq. (\ref{equa7}) can be performed perturbatively in this parameter, yielding 
\begin{equation}
z_{wb}=\left(3\over 2 \kappa \right)^{1/2}\frac{\alpha^
     {3}\,\left(4\,{\sqrt{1 - \chi}} + \chi-1
      \right) }{{\left( {\sqrt{3}} -
        \alpha \right) }^2\,
    \left({\sqrt{1 - \chi}} + \chi -1\right) },
\label{equa8}
\end{equation}
where $\alpha \equiv {(1 + 2\,{\sqrt{1 - \chi}} - \chi})^{1/2}$. 

\begin{figure} [ht]
\includegraphics[scale=0.35]{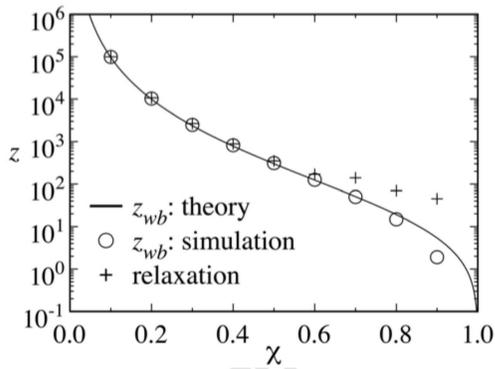}% Here is how to import
%EPS art
\caption{Comparison of the predicted time of the first wave breaking Eq.(\ref{equa8}) (solid line), 
with the result of  dynamics simulations (circles). Time is measured in units of $\kappa^{-1/2}$. 
A very good agreement between the theory and the simulations extends all the way to $\chi=0.8$.
Crosses show the relaxation time --- when emittance saturates. Note that up to $\chi=0.6$ the
wave breaking time and the relaxation times are almost identical, see the discussion in the text. \label{fig3.eps}}
\end{figure}

For small values of $\chi$ the first wave breaking will happen after a long transitory period; 
$z_{wb}\approx 81 \sqrt{2}/\chi^{3}$ as $\chi \rightarrow 0$.  
The subsequent breaks, however, occur on a much shorter
time scale  $\sim 1/\omega_0$, as can be seen from Fig. \ref{fig1.eps}.   Therefore,
$z_{wb}$ should also give us a good estimate of the relaxation time for the entire dynamics.  
In Fig. \ref{fig3.eps} we compare the wave breaking time obtained using the N particle dynamics 
simulation described above with $z_{wb}$ given by Eq. (\ref{equa8}). 
%This is done in our last figure, Fig. \ref{fig3.eps} where for the simulation we simply follow the 
%dynamics in phase-space $r,v$ until the time of the first break. 
The figure reveals an amazingly good 
agreement between the two results. 
In the same plot we also show the relaxation time --- defined as the time when the 
emittance first reaches its plateau value, see Fig. \ref{fig2.eps} (f). As expected, for smaller values of 
$\chi$ the time of relaxation follows closely the wave breaking time.  This is because 
the phase mixing and
the jet production occur on a much shorter time scale, $\sim 1/\omega_0$, than $z_{wb}$.   For larger
values of $\chi$ the two time scales, however, become comparable.  This results in a deviation between the
two data sets --- circles and crosses --- observed in Fig. \ref{fig3.eps}.  

To summarize, we have investigated the dynamics of space-charge dominated  
beams~\cite{davidson01} 
with inhomogeneous density profiles.  Using Lagrangian coordinates, 
we were able to derive a very accurate analytical 
expression Eq. (\ref{equa1p5}) which describes very well the dynamics of beam particles, 
up to the wave breaking time.  
The fluid picture loses its validity when the propagating wave fronts result in
a singular build up of density.  At this point the crest of the 
propagating wave will break off producing an azimuthally symmetric jet of 
particles accelerated by the wave front.
This process will repeat itself many times leading to a final stationary state in which
the beam separates into a cold core surrounded by a halo of highly energetic particles.
The theory presented in this paper allows to calculate precisely the time at which the wave breaking
will take place.   It also gives a very good estimate for the time of relaxation to the 
final stationary state.  
%Wave breaking will occur in any beam system containing a finite amount of inhomogeneity. 
Unlike other systems in which the wave breaking occurs only when thresholds on driving fields are 
exceeded~\cite{rosen02}, inhomogeneous beams are found to be always unstable~\cite{hof83} and the wave 
breaking is unavoidable. 
%Wave breaking produces particle jets that can be responsible for the appearance of kinetic effects in the 
%system. Kinetic effects make the connection between coordinates and velocities multi-valued which breaks 
%down the simple fluid picture for the whole dynamics. The presence of kinetic effects and jets is somewhat 
%equivalent to heating, whereby coherent fluctuations of density are gradually converted into microscopic degrees of 
%freedom associated to beam emittance. 

Wave breaking is not the only mechanism which leads to the relaxation of initially non-stationary intense 
particle beams. It is well known
that oscillations of mismatched envelopes can be damped by the Gluckstern resonances. However, in practice,
inhomogeneities are much harder to suppress than the envelope mismatches \cite{lund05}.  In these cases 
the wave breaking described in the present letter 
will be the dominant mechanism by which a system reaches its final stationary state. 

The theory presented here 
describes beams with vanishing initial emittances.  One example of this are the crystalline beams for which the 
initial emittance is suppressed by a series of 
dissipative cooling procedures \cite{oka02}.  We expect, however,  that the theory will remain valid 
also for beams 
of initial finite emittances as long as the thermal length is small,
$v_{th} z_{wb} = \varepsilon_{th} z_{wb}/2 r_b < A \sim \chi r_b$, 
where $v_{th}=\sqrt{<v^2>}$ is the characteristic thermal velocity,
$A$ is the characteristic amplitude of particle oscillations, and the value of the wave breaking time is
given in Fig. (\ref{fig3.eps}). Since in general $z_{wb} \gg 1/\omega_0$ and $\chi < 1$, the condition requires 
that the thermal velocity be smaller than the macroscopic velocity $\omega_0 A$ \cite{mori88} and 
that the beam be space charge dominated $\varepsilon \ll K/\sqrt{\kappa}$.

This work is supported by CNPq and FAPERGS, Brazil, and by the Air Force Office of Scientific Research 
(AFOSR), USA, under the grant FA9550-06-1-0345.


\begin{thebibliography}{99}
%\bibliography{}

\bibitem{reiser91} A. Cuchetti, M. Reiser, and T. Wangler, in {\it Proceedings of
the Invited Papers, $14^{th}$ Particle Accelerator Conference}, San Francisco, USA, 
1991, edited by L. Lizama and J. Chew (IEEE, New York, 1991), Vol. {\bf 1}, p. 251.
%; M. Reiser, J. App. Phys. {\bf 70}, 1919 (1991).

\bibitem{gluck94} R.L. Gluckstern, Phys. Rev. Lett. {\bf 73}, 1247 (1994).

%\bibitem{lili91} A. J. Lichtenberg and M. A. Lieberman, {\it Regular
%and Stochastic Motion} (Springer-Verlag, New York, 1992), p. 115.

\bibitem{pak94} R. Pakter, G. Corso, T.S. Caetano, D. Dillenburg, and F.B. Rizzato, 
Phys. Plasmas, {\bf 12}, 4099 (1994).

%\bibitem{ibric04} M. Roberto, E.C. da Silva, I.L. Caldas, and R.L. Viana, 
%Phys. Plasmas {\bf 11}, 214 (2004).

\bibitem{pak97} R. Pakter, S.R. Lopes, and R.L. Viana, Physica D {\bf 110}, 277 (1997). 

%AQUI

\bibitem{gluc70}R. L. Gluckstern, {\it Proc. National Accelerator Lab. Linear Accelerator Conf., 
Batavia, IL, Sep. 1970}, p. 811.

\bibitem{lund98}S. M. Lund and R. C. Davidson,  Phys. Plasmas, {\bf 5}, 3028 (1998). 

\bibitem{bern99} S. Bernal, R.A. Kishek, M. Reiser, and I. Haber, Phys. Rev. Lett. {\bf 82}, 4002 (1999).

\bibitem{lund05} S.M. Lund, D.P. Grote, and R.C. Davidson, Nuc. Instr. and Meth. A {\bf 544}, 472 (2005); Y. Fink, C. Chen and W.P. Marable, Phys. Rev. E {\bf 55}, 7557 (1997).

%\bibitem{row90} E. Infeld and G. Rowlands, {\it Nonlinear Waves, Solitons and Chaos} (Cambridge University 
%Press, Cambridge, 1990)

\bibitem{ikegami97} H. Okamoto and M. Ikegami, Phys. Rev. E {\bf 55}, 4694 (1997).

\bibitem{davidson01} M. Reiser, {\it Theory and Design of Charged Particle Beams}, (Wiley-Interscience, 1994); R.C. Davidson and H. Qin, {\it Physics of Intense Charged
Particle Beams in High Energy Accelerators} (World Scientific, Singapore,
2001). 
%pgs. 269 (Energy Conservation), and 274 (Envelope Equation).

\bibitem{pak01} R. Pakter and F.B. Rizzato, Phys. Rev. Lett. {\bf 87}, (2001).

\bibitem{bru95} D. Bruhwiler and Y.K. Batygin, in {\it Proceedings of the PAC 1995, Dallas, TX} 
(IEEE, Piscataway, NJ, 1995), p. 3254.

\bibitem{morr98} P.J. Morrison, Rev. Mod. Phys. {\bf 70}, 467 (1998).

\bibitem{brhe} W. Braun and K. Hepp, Comm. Math. Phys.  {\bf 56}, 101 (1977).

\bibitem{levin02} Y. Levin, Rep. Prog. Phys. {\bf 65}, 1577 (2002).

\bibitem{rosen02} R.J. England, J.B. Rosenzweig, and N. Barov, Phys. Rev. E {\bf 66}, 016501 
(2002); T. Ohkubo, S.V. Bulanov, A.G. Zhidkov, T. Esirkepov, J. Koga, 
M. Uesaka, and T. Tajima, Phys. Plasmas {\bf 13}, 103101 (2006).

%\bibitem{bu06} T. Ohkubo, S.V. Bulanov, A.G. Zhidkov, T. Esirkepov, J. Koga, 
%M. Uesaka, and T. Tajima, Phys. Plasmas {\bf 13}, 103101 (2006).

\bibitem{hof83} I. Hofmann, L.J. Laslett, L. Smith, and I.
Haber, Part. Accel., {\bf 13}, 145 (1983).

%AQUI
\bibitem{oka02} H. Okamoto, Phys. Plasmas {\bf 9}, 322 (2002).

\bibitem{mori88} T. Katsouleas and W.B. Mori, Phys. Rev. Lett. {\bf 61}, 90 (1988).

%\bibitem{lapostolle71} P. M. Lapostolle, IEEE Trans. Nucl. Sci.,
%{\bf NS-18}, 1101 (1971); F. J. Sacherer, IEEE Trans. Nucl. Sci., {\bf NS-18}, 1105
%(1971).

%\bibitem{hess04} M. Hess and C. Chen, Phys. Rev. ST Accel. Beams {\bf 7}, 092002
%(2004).

%\bibitem{struck84} C.J. Struckmeier and
%M.  Reiser, Part. Accel., {\bf 14}, 227, (1984).

%\bibitem{pakter2000} R. Pakter and C. Chen, Phys. Rev. E {\bf 62}, 2789 (2000).

%\bibitem{lau98} {\it Special issue on high-power microwave
%5generation}, edited by E. Schamiloglu and Y.Y. Lau, IEEE
%Trans. Plasma Sci. {\bf PS-26} (1998);
%{\it Space-charge dominated beams and applications
%of high-brightness beams}, edited by S.Y. Lee, AIP Conf.
%Proc. {\bf 377} (AIP, New York, 1996).

%\bibitem{jameson} R.A. Jameson, AIP Conf. Proc. {\bf 279},
%969 (1993).

%\bibitem{agee98} F.J. Agee, IEEE Trans. Plasma Sci., {\bf 26},
%235 (1998).

%\bibitem{struck84} C.J. Struckmeier and
%M.  Reiser, Part. Accel., {\bf 14}, 227, (1984);
%S.M. Lund and B. Bukh, Phys. Rev. ST Accel. Beams {\bf 7}, 024801 (2004);
%J.S. Moraes, R. Pakter, F.B. Rizzato, and C. Chen, Phys. Scripta
%{\bf T107}, 145 (2004).

%\bibitem{pakter01} R. Pakter and F.B. Rizzato, Phys. Rev. Lett.
%{\bf 87}, 044801 (2001); Phys. Rev. E, {\bf 65},
%056503 (2002).

%\bibitem{chen94a} C. Chen and R.C. Davidson, Phys. Rev. Lett.,
%{\bf 72}, 2195 (1994); R.L. Gluckstern, Phys. Rev.
%Lett. {\bf 73} 1247 (1994); A. Riabko, M. Ellison, X. Kang,
%S. Y. Lee, D. Li, J. Y. Liu, X. Pei, and L. Wang, Phys. Rev. E,
%{\bf 51}, 3529 (1995); R.L. Gluckstern, W.-H. Cheng, and H. Ye, Phys. Rev.
%Lett. {\bf 75} 2835 (1995).

%\bibitem{pakter00} R. Pakter and C. Chen, IEEE Trans. Plasma Sci.,
%{\bf 28}, 502 (2000).

%\bibitem{chen03} J. Zhou, B.L. Qian, and C. Chen, Phys. Plasmas,
%{\bf 10}, 4203 (2003).

%\bibitem{kv59} I.M. Kapchinskij and V.V. Vladimirskij, in {\it
%Proceedings of the International Conference on High Energy Accelerators}
%(CERN, Geneva, 1959), p. 274.

\end{thebibliography}
\end{document}